\documentstyle[epsf]{elsart}
\begin{document}

\epsfverbosetrue

\begin{frontmatter}

\title{Probing a Nucleon Spin Structure at TESLA\\ by the Real
Polarized Gamma Beam }

\author[protvino]{S. Alekhin}
\author[protvino]{V. Borodulin}
\address[protvino]{Institute for High Energy Physics, Protvino, 
Russia}
\author[ankara]{A. \c{C}elikel}
\author[ankara]{M. Kantar}
\author[ankara]{S. Sultansoy\thanksref{baku}}
\address[ankara]{Ankara University, Dept.of Physics, Ankara, Turkey }
\thanks[baku]{On leave from Institute of Physics,
Baku, Azerbaijan}

\begin{abstract}
The recent proposals concerning the usage of 
the real polarized
gamma beam, obtained by the  Compton backscattering
of the laser photons off the electron beams from
either the linear or circular accelerators were considered.
The heavy quark photoproduction process giving a unique
opportunity to measure polarized gluon distribution
was investigated.
\begin{keyword}
Asymmetry, spin, charm quark, laser, high energy $\gamma$-beam
\end{keyword}

\end{abstract}
\end{frontmatter}

\section{\bf Introduction}

A spin crisis arose when the first determination of 
\begin{eqnarray}
\Delta \Sigma =\Delta u+\Delta d+\Delta s  \nonumber
\end{eqnarray}
was found to be much smaller than expected \cite{EMC}, where 
\begin{eqnarray}
\Delta f\equiv \int dx\Delta f(x,Q^2),  \nonumber
\end{eqnarray}
and $f(x,Q^2)$ are the polarized quark spin distribution functions. The
recent value of the world average for $\Delta \Sigma $ is approximately $%
0.3\pm 0.06$. It is much smaller than the relativistic quark model
prediction of 0.6 \cite{Ellis}. Among the number of explanations of the
EMC
results on the longitudinally polarized proton structure function, that of
the perturbative QCD approach occupies a prominent position and opens a
new
domain of tests \cite{Efremov}. It has been shown that in perturbative QCD
predictions of the quark-parton model concerning the singlet axial charge
contribution to $I_1^p$ need to be modified because of the $\gamma _5$
anomaly of the flavour singlet current $J_\mu ^5$. In this approach the
main
question is the size of the polarized gluon distribution. But there is a
great uncertainty in theoretical estimates on the magnitude and $x$%
--dependence of $\Delta G(x)$, especially in the moderate $x$--region
$x>0.1$%
. One calculation, performed in the framework of MIT Bag Model, predicts
negative $\Delta G(x)$ value, thus even sharpening the problem \cite{MIT}.

In order to measure gluon contribution to the nucleon spin, we must select
a
process where the familiar lowest order graphs of deep inelastic
scattering
from a single quark are suppressed. Analysis of the experimental data
shows
that there is no significant charm content in the nucleons
\cite{charmsea}.
Then one may hope to determine polarized gluon effects in the process of
the
charmed quark photoproduction. Due to the comparatively large mass of the
$c$%
--quark the leading mechanism for obtaining a $c \bar c$- pair in the
final
state is the hard photon-gluon fusion (PGF) process 
\begin{eqnarray}
\gamma + g \rightarrow c + \bar c
\end{eqnarray}
and produced $c\bar c$-- pair may form $J/\psi$ meson or fragment
separately
into open charm states.

A rich spin physics programme was proposed by COMPASS collaboration at
CERN 
\cite{Compass}. A large part of it is devoted to measurement of gluon
polarization by tagging $D^0,\;D^{*+}$-mesons produced in the $\gamma
^{*}N$
collisions. Produced $D^0,\;D^{*+}$-mesons will be reconstructed from
their
two-and three-body decays into hadrons. It permits to impose constraints
on
the invariant masses of $K\;\pi \;\pi ,\;K\;\pi ,..$ subsystems, which
effectively rejects a background, especially in the case of $D^{*+}$%
-tagging. A statistical precision on the polarization asymmetry
measurement
is expected to be about 0.05.

A proposal has been made \cite{asympropos} to determine $\Delta G(x)$ from
the measurement of the asymmetry in the charm photoproduction process in
the
scattering of polarized real photons off polarized fixed target. In this
paper we consider a possibility to realize this experiment at the TESLA
machine \cite{tesla}, and discuss briefly an opportunity to perform it at
SLAC and LEP2.

\newpage

\section{The Real Gamma Beam}

\hspace*{30mm} {\bf 2.1 Linear Accelerator} \hspace*{10mm}

The scheme of the proposed experiment looks as follows. Circularly
polarized
laser beam with photon energy $\omega _0=3.3\;eV$ ( Cu15 laser) is
scattered
off the $250\;(50)\;GeV$-electrons provided by TESLA (SLAC) \cite{Ginzb}.
Throughout this subsection the numbers in parentheses refer to the SLAC
gamma beam parameters. The Compton back-scattered photon beam has a
spectrum 
$\gamma (y)$ 
\begin{eqnarray}
\gamma (y) &=&\frac 1N\times \frac{dN}{dy},\;\;\;N=\int dy\frac{dN}{dy}, 
\nonumber \\
\frac{dN}{dy} &=&\frac 1{1-y}+1-y-4r(1-r),  \label{spectr}
\end{eqnarray}
and follows closely to the trajectory of the primary electron beam. Here
the
ratio of the hard photon energy $\omega $ to the electron energy $E_e$
ranges as 
\begin{eqnarray}
y &=&\frac \omega {E_e},\;\;\;0\leq y\leq y_{max}=\frac \kappa {1+\kappa
}\sim 0.927\;(0.717),  \nonumber \\
\kappa &=&\frac{4\omega _0E_e}{m_e^2}\sim 12.638\;(2.528),\;\;r=\frac
y{\kappa (1-y)}.
\end{eqnarray}
Had the laser beam been circularly polarized, the $\gamma $-beam would be
polarized too and the $\gamma $-beam helicity is given by 
\begin{eqnarray}
\xi _2 &=&\frac{B(y)}{N\gamma (y)},  \nonumber \\
B(y) &=&-\lambda _{ph}(2r-1)\Bigl(\frac 1{1-y}+1-y\Bigr).  \nonumber
\end{eqnarray}
where $\lambda _{ph}$ denotes the laser photon helicity. Due to the unique
dependence of both $\gamma $-beam energy and polarization on the
scattering
angle $\theta _\gamma $ between the incident and scattered photon 
\begin{eqnarray}
\theta _\gamma (\omega )\sim \frac{m_e}{E_e}\sqrt{\frac{E_e\;\kappa
}\omega
-(\kappa +1)},  \nonumber
\end{eqnarray}
it is possible to obtain almost monochromatic gamma beam with energy %
\hbox{$0.99 \omega_{max}\leq \omega \leq \omega_{max} $} and polarization
nearly equal to unit by selecting photons with $\theta _\gamma \leq
0.759\;(1.93)\cdot 10^{-6}rad$. Taking the distance between the conversion
region and collimator $100m$, we obtain that this angle corresponds to the
diameter of the selecting slit $d=152\;(386)\mu m$.

In our further estimation on the luminosity of gamma beam scattering off a
polarized fixed target, we shall follow to the method outlined in
\cite{Atag}%
. We determine the optimal number of converted photons $N_\gamma $ by the
requirement to obtain one event per collision 
\begin{eqnarray}
\beta N_\gamma T_n\sigma _{\gamma p}=1,  \label{restr}
\end{eqnarray}
where $\beta $ is the fraction of the photons coming through the slit,
$T_n$
stands for the target nucleon density and $\sigma _{\gamma p}$ is the
total
cross section of gamma-proton collisions. In the case of $1\%$ $\gamma $%
-beam monochromaticity $\beta \sim 4.22\;(2.02)\cdot 10^{-2}$, for
deuterated butanol target with the length about 40 $cm$ the density is $%
T_n=4\cdot 10^{25}cm^{-2}$ and $\sigma _{\gamma p}\sim 100\mu b$ at high
energies, so we get $N_\gamma =5.92\;(12.36)\cdot 10^3$. Because the
number
of electrons in a bunch $N_e=5.15\;(3.5)\cdot 10^{10}$ for TESLA (SLAC),
we
obtain the necessary conversion coefficient 
\begin{eqnarray}
K=\frac{N_\gamma }{N_e}=1.15\;(3.53)\cdot 10^{-7}.  \nonumber
\end{eqnarray}

The number of photons required to provide the scattering of each electron
with the laser photon is defined from 
\begin{eqnarray}
\frac{n_0\sigma _c}{S_{eff}}=1,  \label{s_eff}
\end{eqnarray}
where $n_0$ is the number of photons in a laser pulse, $\sigma
_c=1.1\;(2.59)\cdot 10^{-25}cm^2$ is the total Compton cross section and
$%
S_{eff}$ is the effective area of the photon and electron beams intercept.
Choosing $S_{eff}=4\cdot 10^{-6}cm^2$, one can easily get the laser energy
per pulse 
\begin{eqnarray}
A_0=n_0\;\omega _0\sim 19.2\;(8.152)\;J,  \nonumber
\end{eqnarray}
corresponding to the total electron conversion. Since the optimal number
of
converted photons $N_\gamma $, defined by (\ref{restr}), is much less than
$%
N_e$, we need not the total conversion of electrons and the required pulse
energy is suppressed by the factor of $K$ 
\begin{eqnarray}
A=K\;A_0\sim 2.21\;(2.88)\cdot 10^{-6}\;J,  \nonumber
\end{eqnarray}
which is accessible with modern laser technology.

Finally we shall discuss a choice of the laser pulse frequency. It should
evidently coincide with the linac frequency $f_{pulse}$ when number of the
bunches $n_b$ in the electron beam is equal to unity. If the electron beam
has a multybunch structure, one may either choose the repetition rate of
laser pulses to be equal to $f_{pulse}\cdot n_b$ or use a mirror system
\cite
{mirror}, which evidently decreases the laser pulse frequency. Since the
detailed discussion of an experimental setup is beyond the scope of our
paper, we choose for simplicity the laser pulse frequency equal to $%
f_{pulse}\cdot n_b$.

For the linear accelerators the integrated luminosity per a year of
operation takes the form 
\begin{eqnarray}
{\ L}^{int}_{linac} = 10^7 \cdot \; f_{rep} \; \beta \; N_e \; K \; T_n,
\\
f_{rep} = f_{pulse}\cdot n_b,  \nonumber
\end{eqnarray}
where $f_{rep}$ is the collision frequency. Taking into account the
Eq.(\ref
{restr}), one can estimate $L^{int}_{linac}$ as follows 
\begin{eqnarray}
{\ L}^{int}_{linac} = 10^7 \cdot \frac{f_{rep}}{\sigma_{\gamma p}} \sim
0.8\; (0.012)\; fb^{-1}.
\end{eqnarray}

\hspace*{30mm} {\bf 2.2 Circular Accelerator} \hspace*{10mm}

Now we briefly consider the possibility to obtain the real gamma beam at a
circular electron accelerator, say at LEP2. All calculations are done in
analogy with the TESLA case, except a few details. First of all, $f_{rep}$
now takes the form 
\begin{eqnarray}
f_{rep}=f_{pulse}\cdot n_b=\frac c{2\pi R}\cdot n_b=44\;980\;Hz,
\nonumber
\end{eqnarray}
where the number of electron bunches $n_b=4$, $c$ is the speed of light,
$R$
is the ring radius and $f_{pulse}$ is the circulation frequency in this
case. It is evident that for the circular accelerator $f_{rep}$ should
coincide with frequency of the laser pulses. The next remark concerns the
Eq.(\ref{s_eff}). We choose the effective area of the photon and electron
beams intercept to be $S_{eff}\sim S_e=2\cdot 10^{-4}cm^2$, where $S_e$ is
the transverse area of the LEP2 electron beam, and the total Compton cross
section $\sigma _c=1.86\cdot 10^{-25}cm^2$ at the LEP2 energy. And
finally,
we should take into account the effects of beam mean life time ( we did
not
consider these effects earlier, since every electron bunch provided by the
linear accelerator could be used only once ).

\begin{table}[tbp]
\caption{The parameters of the LEP2, SLAC and TESLA electron beams. Here
$%
f_{pulse}$ is the pulse frequency, $H$ and $V$ are the horizontal and
vertical beam radius correspondingly, $N_e$ is the number of electrons per
bunch and $n_b$ is the number of bunches.}
\begin{tabular}[2]{|c|c|c|c|c|c|c|}
\hline
&  &  &  &  &  &  \\ 
& $E_e$ (GeV) & $N_e\;( 10^{10})$ & \hspace*{5mm} $n_b$\hspace*{5mm} & $%
f_{pulse}$(Hz) & H ($\mu m$ ) & V ( $\mu m$ ) \\ \hline
&  &  &  &  &  &  \\ 
LEP2 \hspace*{-1mm} & 100 & 40 & 4 & $11.25\cdot 10^3$ & 200 & 8 \\ \hline
&  &  &  &  &  &  \\ 
SLAC \hspace*{-1mm} & 50 & 3.5 & 1 & 120 & 2.1 & 0.6 \\ \hline
&  &  &  &  &  &  \\ 
TESLA \hspace*{-1mm} & 250 & 5.15 & 800 & 10 & 0.64 & $0.1$ \\ \hline
\end{tabular}
\vspace*{1mm}
\end{table}
In each collision about $K\;N_e$ electrons are scattered. Since every
bunch
exercise the collision with the frequency $f_{pulse}=c/2\pi R\sim
1.12\cdot
10^4Hz$, the mean lifetime of the beam $\tau _b$ may be estimated as 
\begin{eqnarray}
\tau _b=\frac{\log (1-\delta )}{\log (1- K)}\cdot \frac 1{f_{pulse}}\sim
400\;s,  \nonumber
\end{eqnarray}
where $\delta $ denotes the maximal fraction of electron loss permitted by
beam dynamics. In what follows we choose $\delta =0.1$, then the number of
collisions each bunch exercises during one cycle, is equal to 
\begin{eqnarray}
l=\frac{\log (1-\delta )}{\log (1-K)}\sim 4.46\cdot 10^6,  \nonumber
\end{eqnarray}
and the mean number of electrons in a bunch $\hat{N_e}$ is given by 
\begin{eqnarray}
\hat{N_e}=(1-\frac{l}{2}k)N_e,  \nonumber
\end{eqnarray}
where $N_e$ denotes the initial number of electrons in a bunch.

For the ring accelerators the integrated luminosity per year of operation
takes the form 
\begin{eqnarray}
{L}_{ring}^{int}=10^7\cdot \frac{\tau _b}{\tau _a+\tau _b+\tau _f}%
\;f_{rep}\;\beta \;\hat{N_e}\;K\;T_n,
\end{eqnarray}
where $\tau _a$ is an acceleration time, $\tau _f$ is a filling time.
Taking
into account the Eq.(\ref{restr}), one can estimate $L_{ring}^{int}$ as
follows 
\begin{eqnarray}
{L}_{ring}^{int}=10^7\cdot \frac{\tau _b}{\tau _a+\tau _b+\tau _f}f_{rep}%
\frac{\hat{N_e}}{N_e}\frac 1{\sigma _{\gamma p}}\sim 0.27fb^{-1}.
\end{eqnarray}
The electron beam parameters, we have used in the estimations, are
contained
in Tabl. 1. while the $\gamma $-beam characteristics are given in Tabl. 2.

\begin{table}[tbp]
\caption{ The real gamma beam parameters at LEP2, SLAC and TESLA. Here $%
y_{max}$ is the ratio of the gamma beam maximal energy to the electron
energy, $\theta _\gamma $ is the opening angle of the collimator, $\beta $
denotes the fraction of photons coming through the slit, $N_\gamma $ is
the
total number of converted photons, K is the conversion coefficient and A
is
the laser energy per pulse. }
\begin{tabular}{|c|c|c|c|c|c|c|c|}
\hline
&  &  &  &  &  &  &  \\ 
& $\hspace*{3mm} y_{max} \hspace*{3mm}$ & $\theta_{\gamma} $ &
$\hspace*{5mm}%
\beta \hspace*{5mm}$ & \hspace*{3mm}$N_{\gamma}$ \hspace*{3mm} & $%
\hspace*{5mm} K \hspace*{5mm}$ & $A $ & $L^{int}$ \\ 
&  & \hspace*{-1mm} $(10^{-6} rad)$\hspace*{-1mm} &  & $(10^{4})$ & $%
(10^{-7})$ & $(10^{-6} J)$ & $(fb^{-1})$ \\ \hline
&  &  &  &  &  &  &  \\ 
LEP2\hspace*{-1mm} & 0.835 & 1.26 & 0.0265 & 0.945 & 0.236 & 13.4 & 0.27
\\ 
\hline
&  &  &  &  &  &  &  \\ 
SLAC\hspace*{-1mm} & 0.717 & 1.93 & 0.0202 & 1.236 & 3.53 & 2.88 & 0.012
\\ 
\hline
&  &  &  &  &  &  &  \\ 
TESLA\hspace*{-1mm} & 0.927 & 0.759 & 0.0422 & 0.592 & 1.15 & 2.21 & 0.8
\\ 
\hline
\end{tabular}
\vspace*{1mm}
\end{table}

\vspace*{1mm}

Thus the luminosity at a probable SLAC experiment is about one order of
magnitude smaller compared with LEP2 case due to a lower collision
frequency. Choosing a more powerful laser one can reach almost the same
luminosity at SLAC, but the price will be too high: more than one hundred
events per collision, which reduces significantly the range of the
physical
phenomena accessible for investigations. The luminosity at TESLA even
exceeds the LEP2, since the TESLA project will operate with multybunch
trains.

\section{ The Heavy Quark Photoproduction}

The differential cross section of the photon-gluon fusion looks as follows 
\cite{difcrossec} 
\begin{eqnarray}
\frac{d \hat\sigma}{d\hat t}= \frac{d \hat\sigma^n}{d\hat t} + \lambda_g
\xi_2 \frac{d \delta\hat\sigma}{d\hat t},  \nonumber
\end{eqnarray}
where $\lambda_g$ and $\xi_2$ denote the gluon and photon helicities. The
spin averaged and polarized asymmetry distributions in the LO QCD take the
form, corespondingly: 
\begin{eqnarray}
\frac{d \hat\sigma^n}{d\hat t}= \frac{\pi\alpha\alpha_s (s) e^2_q}{\hat
s^2} %
\Biggl( \frac{4 m^2 \hat s}{(\hat t - m^2)(\hat u - m^2)} + \frac{\hat u -
m^2}{\hat t - m^2} + \frac{\hat t - m^2}{\hat u - m^2}  \nonumber \\
- \frac{4 m^4 \hat s^2}{{(\hat t - m^2)}^2{(\hat u - m^2)}^2} \Biggr).
\end{eqnarray}
\begin{eqnarray}
\frac{d \delta \hat\sigma}{d\hat t}= \frac{\pi\alpha\alpha_s (s) e^2_q}{%
2\;\hat s^2} \Biggl( \Bigl(\frac{\hat u - m^2}{\hat t - m^2} + \frac{\hat
t
- m^2}{\hat u - m^2} \Bigr)^2  \nonumber \\
- \hat s (\hat s - 4\;m^2) \Bigl( \frac{1}{(\hat t - m^2)^2} + +
\frac{1}{%
(\hat u - m^2)^2} \Bigr) \Biggr),
\end{eqnarray}
Here $\hat s, \hat t, \hat u$ are the invariant variables of the
subprocess, 
$\alpha,\; \alpha_s (s)$ are the fine and strong coupling constants
respectively and $e_q$ denotes the c-quark charge.

\input epsf 
\begin{figure}[tbp]
\hbox{\hspace*{10pt}
\epsfxsize=165pt \epsfbox{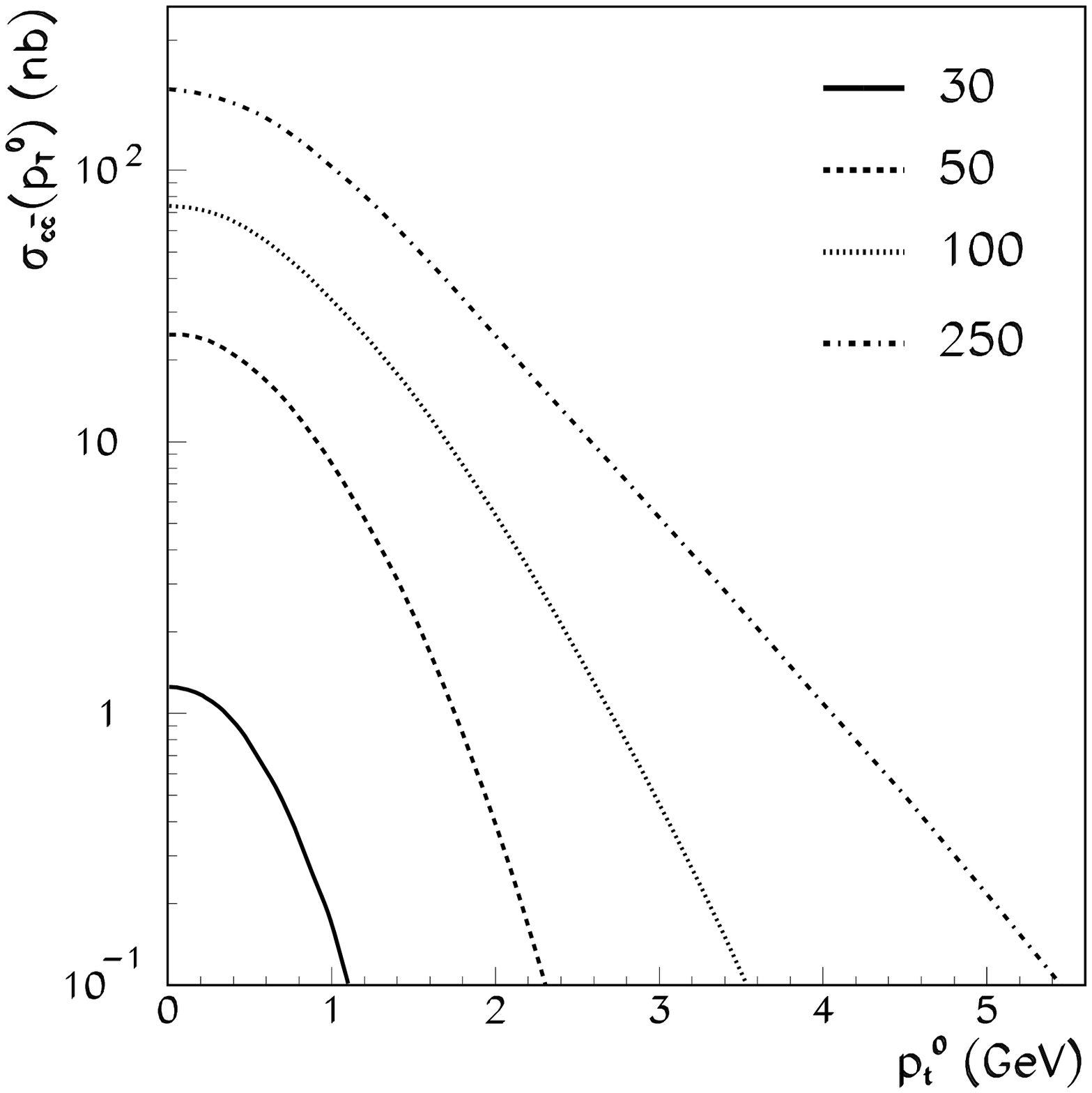}
\hspace*{10pt}
\epsfxsize=165pt \epsfbox{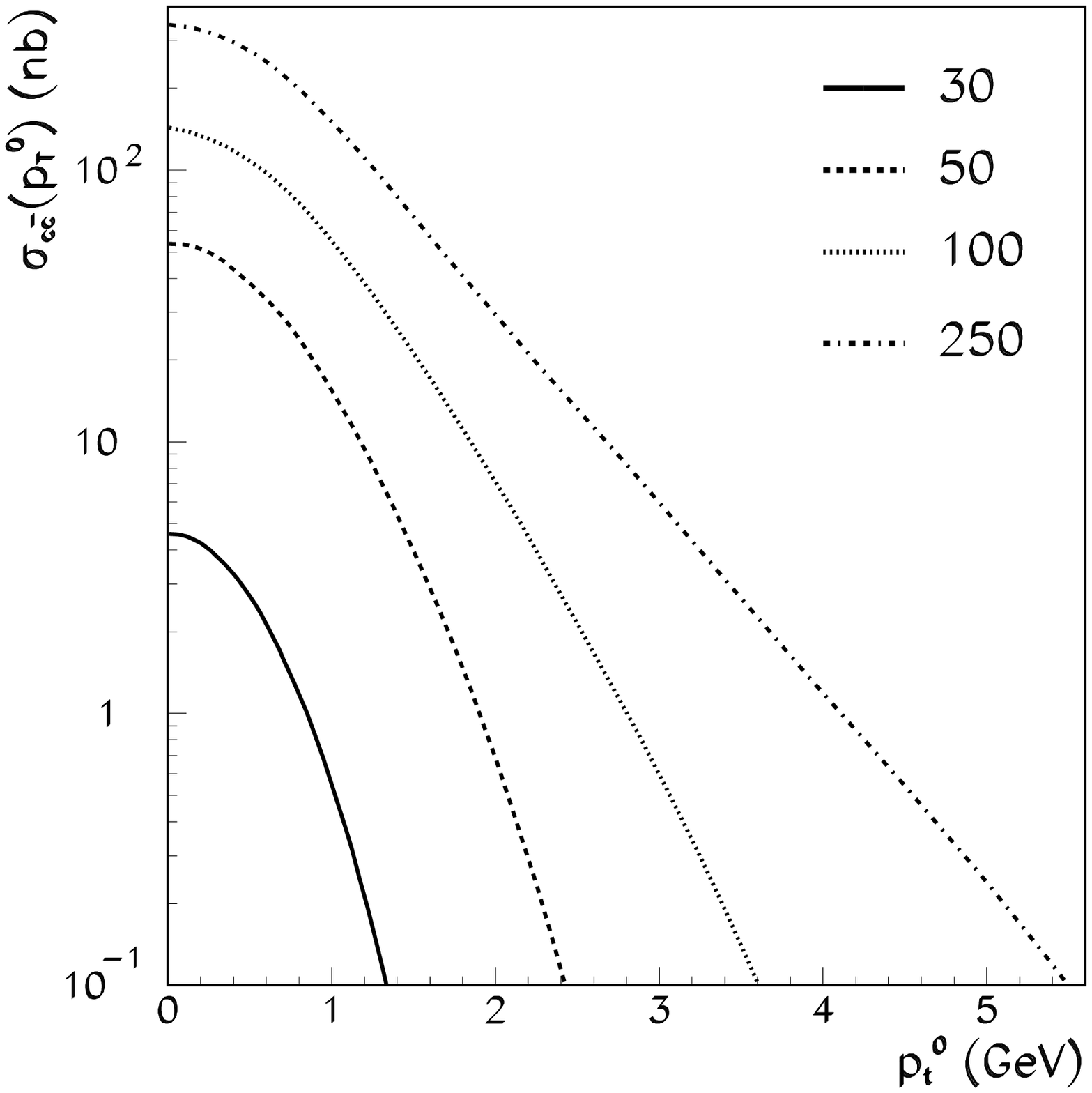}
} \vspace*{-10pt} \hspace*{32mm} ({\small a}) \hspace*{58mm} ({\small b})
\hspace*{8mm}
\par
\vspace*{1mm}
\caption{The distribution of charm production $\sigma_{c\bar c} (p_t\geq
p_t^0)$ versus $p^0_t$ for different electron energies: $E_e = 30\;GeV$ -
solid line, $E_e = 50\;GeV$ - dashed line, $E_e = 100\;GeV$ - dotted
curve, $%
E_e = 250\;GeV$ - dashed-dotted line. The c-quark mass is $m_c =
1.5\;GeV/c^2 $ (a), $m_c = 1.3\;GeV/c^2$ (b). }
\end{figure}

The produced $c\bar{c}$ pairs can form $J/\psi $ or fragment into
$D,\;D^{*}$%
-mesons. In what follows we shall consider only the open charm production,
because this process is more transparent from theoretical point of view.
In
addition, the open charm has an advantage over $J/\psi $ production,
because
its cross-section is at least ten times larger for attainable photon
energies.

\input epsf 
\begin{figure}[tbp]
\hbox{\hspace*{10pt}
\epsfxsize=165pt \epsfbox{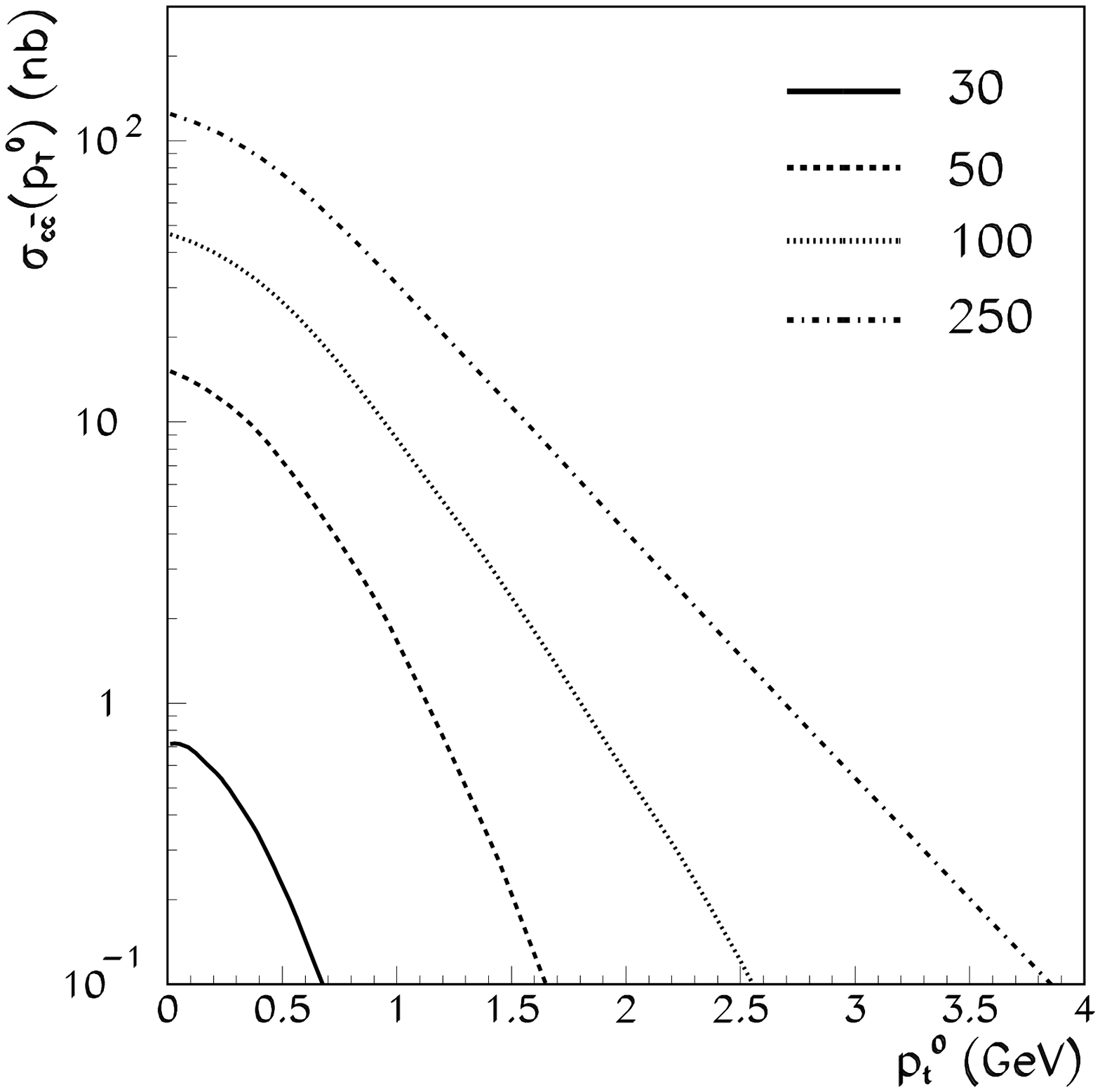}
\hspace*{10pt}
\epsfxsize=165pt \epsfbox{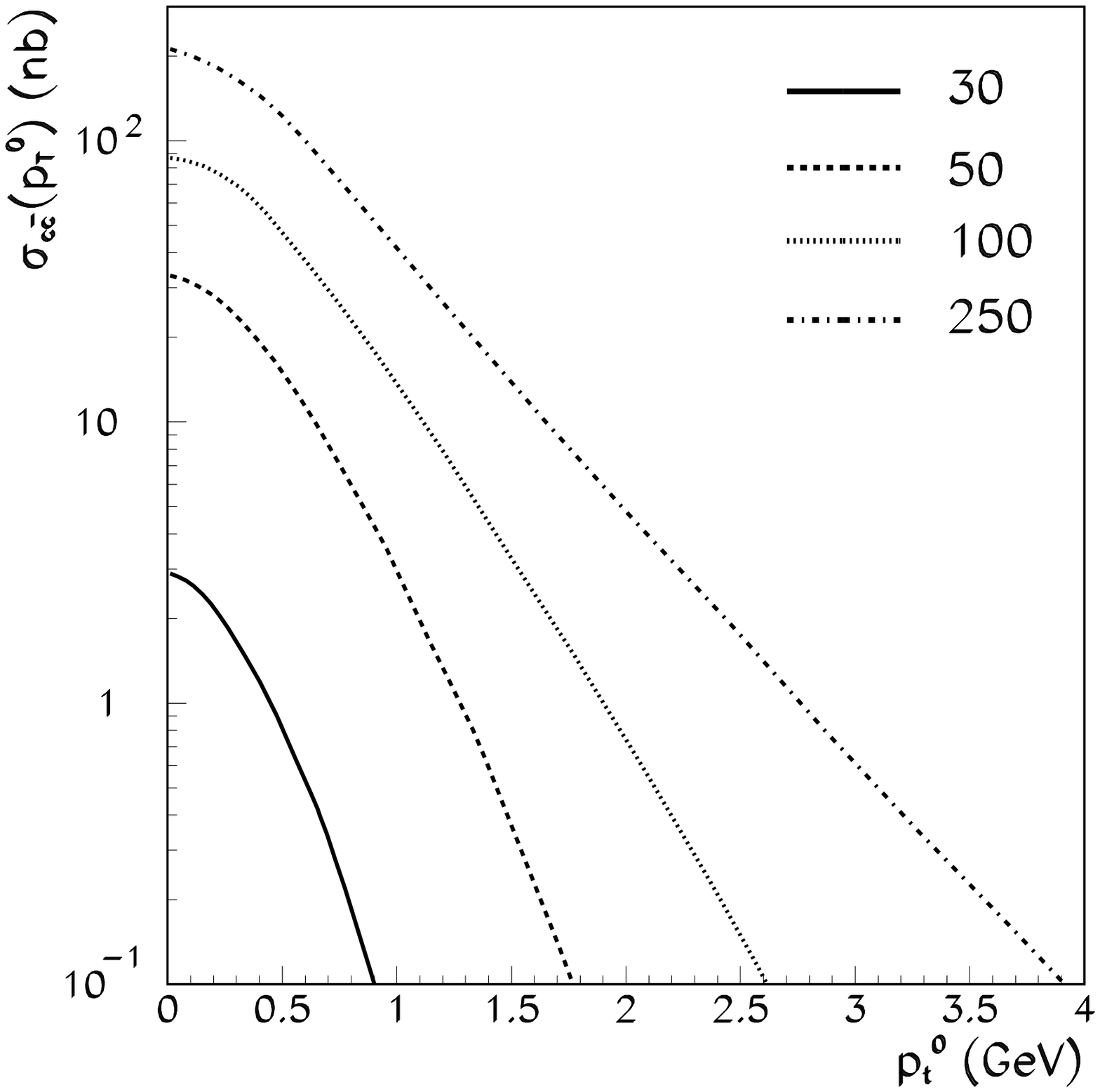}
} \vspace*{-10pt} \hspace*{34mm} ({\small a}) \hspace*{57mm} ({\small b})
\hspace*{6mm}
\par
\vspace*{1mm}
\caption{The distribution of the inclusive $D^*$-meson production $%
\sigma_{D^*\;X} (p_t\geq p_t^0)$ versus $p^0_t$ for different electron
energies: $E_e = 30\;GeV$ - solid line, $E_e = 50\;GeV$ - dashed line,
$E_e
= 100\;GeV$ - dotted curve, $E_e = 250\;GeV$ - dashed-dotted line. The
c-quark mass is $m_c = 1.5\;GeV/c^2$ (a), $m_c = 1.3\;GeV/c^2$ (b). }
\end{figure}

The accumulated $p_t$--distribution of charmed quarks 
\begin{eqnarray}
\sigma _{c\bar{c}}(p_t^0)=\int_{_{_{p_t\geq p_t^0}}}\hspace*{-3mm}dp_t\int 
\frac{d\hat{s}}s\;\frac{d\hat{\sigma}^n}{dp_t}\;G(x,Q^2)
\end{eqnarray}
versus $p_t^0$ is shown in Fig.1, and Fig. 2 presents the same dependence
for inclusive $D^{*}$-meson production 
\begin{eqnarray}
\sigma _{D^{*}}(p_t^0)=\sigma _{D^{*+}}(p_t^0)+\sigma _{D^{*0}}(p_t^0). 
\nonumber
\end{eqnarray}

\input epsf 
\begin{figure}[tbp]
\hbox{\hspace*{72pt}
\epsfxsize=220pt \epsfbox{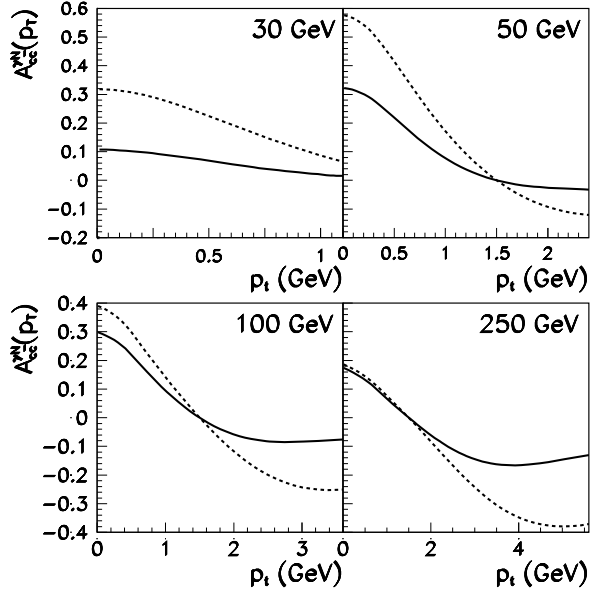}
}
\par
\vspace*{1mm}
\caption{The differential asymmetry $A_{c\bar c}^{\gamma N} (p_t)$ of the
c-quark production versus $p_t$ for the c-quark mass equal to
$1.5\;GeV/c^2$%
. The solid (dashed) line corresponds to the set A (B) of polarized gluon
density [9]. }
\end{figure}

In our estimates we have used the parametrizations of both polarized and
unpolarized gluon densities from \cite{Gehrmann}, $c$--quark fragmentation
functions were taken from \cite{Peterson}, $\alpha _s$ from the global
analysis of DIS data \cite{dis}. A gamma beam spectrum was chosen as
monochromatic with energy $E_\gamma =\kappa /(1+\kappa )E_e$ for $E_e\geq
50\;GeV$, while for $E_e=30\;GeV$ the smeared-out spectrum, given by (\ref
{spectr}), was used. Numerical integration was performed with the help of
adaptive integration code \cite{numercode}. All distributions have similar
form and decrease exponentially with $p_t^0$ rise when transverse momentum
exceeds $1\;GeV/c$. One concludes that the estimated production rate of $%
D^{*}$--mesons with large transverse momentum $p_t\geq 1\;GeV/c$ is
sizeable
and reaches approximately 30 nb (40 nb) at TESLA energy for $c$--quark
mass
equal to $1.5\;(1.3)\;GeV/c^2$ correspondingly.

\input epsf 
\begin{figure}[tbp]
\hbox{\hspace*{75pt}
\epsfxsize=200pt \epsfbox{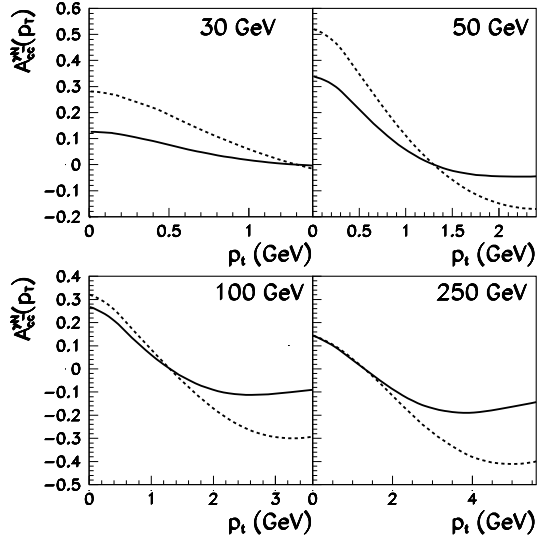}
}
\par
\vspace*{1mm}
\caption{The differential asymmetry $A_{c\bar c}^{\gamma N} (p_t)$ of the
c-quark production versus $p_t$ for the c-quark mass equal to
$1.3\;GeV/c^2$%
. The solid (dashed) line corresponds to the set A (B) of polarized gluon
density [9]. }
\end{figure}

The differential asymmetry $A_{c\bar{c}}^{\gamma N}(p_t)$ of $c$--quark
production versus $p_t$ 
\begin{eqnarray}
A_{c\bar{c}}^{\gamma N}(p_t)=\frac{d\delta \sigma }{dp_t}/\frac{d\sigma
^n}{%
dp_t}
\end{eqnarray}
is shown in Figs.~3--4 for the c-quark mass $1.5\;(1.3)\;GeV/c^2$
correspondingly, and analogous dependences for $D^{*}$--meson production
are
pictured in Figs.~5--6 for the same mass set. Here we have introduced the
following notations 
\begin{eqnarray}
\frac{d\sigma ^n}{dp_t} &=&\int
\frac{d\hat{s}}s\frac{d\hat{\sigma}^n}{dp_t}%
G(x,Q^2),  \nonumber \\
\frac{d\delta \sigma }{dp_t} &=&\int \frac{d\hat{s}}s\frac{d\delta \hat{%
\sigma}}{dp_t}\Delta G(x,Q^2).  \nonumber
\end{eqnarray}

As is already known, the differential asymmetry of $c\bar{c}$ production,
derived in the LO approximation, has a kinematic zero at $p_t=m_c$ and
then
changes a sign. For $D^{*}$--meson production a zero position shifts to
the
lower value of $p_t\sim 1\;GeV/c$ due to fragmentation smearing.
Integration
over total range of $p_t$ evidently decreases the asymmetry, so it is
reasonable to introduce a kinematic cut on the $D^{*}$--meson transverse
momentum, say $p_t\geq 1\;GeV/c$. In this region $A_{D^{*}}^{\gamma
N}(p_t)$
does not change sign, moreover the predicted asymmetry value heavily
depends
on the choice of polarized gluon distribution. It is due to the fact that
the main contribution to production of $D^{*}$-- mesons having large
transverse momentum comes from the region $x\geq 0.1$, where $\Delta
G(x,Q^2) $ is poorely known. At present there are plausible restrictions
on
the polarized gluon density only at $x\leq 0.1$, where all model
predictions
almost coincide. Then the precise measurement of polarized gluon density
at $%
x\geq 0.1$ will surely help to choose a reasonable parametrization of $%
\Delta G(x,Q^2)$.

\input epsf 
\begin{figure}[tbp]
\hbox{\hspace*{75pt}
\epsfxsize=200pt \epsfbox{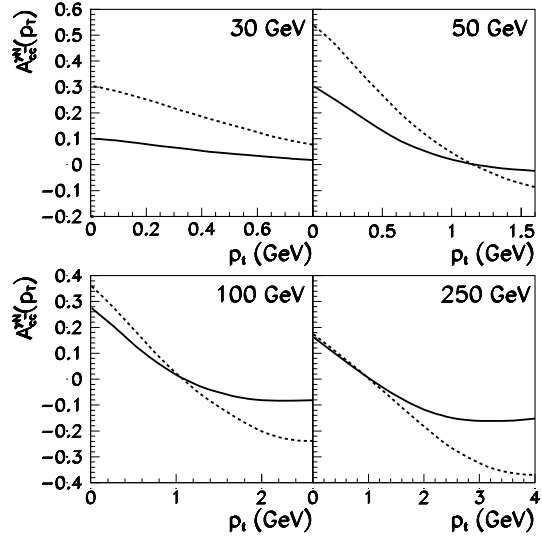}
}
\par
\vspace*{1mm}
\caption{The differential asymmetry $A_{D^*}^{\gamma N} (p_t)$ of the
$D^*$
meson production versus $p_t$ for the c-quark mass equal to
$1.5\;GeV/c^2$.
The solid (dashed) line corresponds to the set A (B) of polarized gluon
density [9]. }
\end{figure}

\input epsf 
\begin{figure}[tbp]
\hbox{\hspace*{72pt}
\epsfxsize=200pt \epsfbox{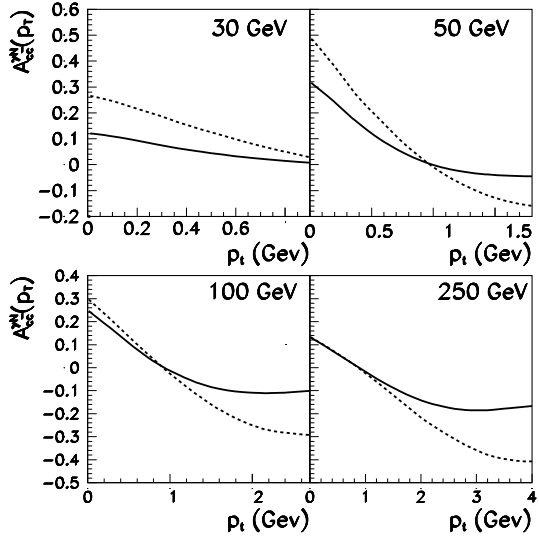}
}
\par
\vspace*{1mm}
\caption{The differential asymmetry $A_{D^*}^{\gamma N} (p_t)$ of the
$D^*$
meson production versus $p_t$ for the c-quark mass equal to
$1.3\;GeV/c^2$.
The solid (dashed) line corresponds to the set A (B) of polarized gluon
density [9]. }
\end{figure}

Let us briefly discuss the next-to-leading (NLO) corrections to the charm
production process. As it was shown in \cite{Stratman}, these corrections
are quite sizeable at SLAC and CERN experiments, so one should introduce
large K-factor to take into account NLO corrections. It is not the case at
TESLA, where NLO correction to the total asymmetry does not exceed
0.3-0.5.
Then one may hope that higher order corrections will also be under control
at possible TESLA experiment. It is not improbable, because all invariants
(taking into account introduced kinematic cut) are sufficiently large at
TESLA energy. The total asymmetry is of the order $10\; \%$ at $\sqrt
s_{\gamma p} \sim 10 \;GeV $ and decreases with energy rise to a few
percent
when $\sqrt s_{\gamma p} \sim 20 \;GeV$ \cite{Stratman}. The introduced
kinematic cut on $D^*$ --meson transverse momentum surely rises this
value,
because the differential asymmetry $A_{D^*}^{\gamma N} (p_t)$ does not
change sign beyond $p_t \sim 1 \;GeV/c$.

\section{ Event Reconstruction}

\hspace*{45mm} {\bf 4.1 $D^*$- tagging} \hspace*{20mm}

One of the best method for $D^*$-- tagging uses the kinematic constraint
of
the decay chain: 
\begin{eqnarray}
D^{*} \rightarrow D \pi \rightarrow (K \pi) \pi.  \label{chain}
\end{eqnarray}
The difference of the invariant masses 
\begin{eqnarray}
\Delta M = m(K \pi\pi) - m(K\pi) = m(D^{*})-m(D) \sim 145 \;MeV/c^2
\end{eqnarray}
is very close to the $\pi$--meson mass, and can be determined with
precision
about $2.5 \;MeV/c^2$ \cite{Compass}, significantly exceeding accuracy of
$%
D^{*}$-meson mass measurement. It permits to reduce substantially the
background to $D^{*+}$-meson production, for example, COMPASS estimates
showed that the background is less than $10\;\%$.

The isospin invariance suggests equal $D^{*+}$ and $D^{*0}$ production
rates, then $\sigma(D^{*+}\;X) \sim 0.5 \cdot \sigma(D^{*}\;X)$. Since the
branching ratio for the decay chain ({\ref{chain}) is 2.6 $\%$, about
$\sim
62.4\cdot 10^{4}$ ($83.2\cdot 10^4$) $D^{*+}$-mesons with $p_t$ exceeding
$%
1\;GeV/c$ will be produced at TESLA energy in a year. Assuming an overall
acceptance to be equal to 0.2 \cite{Compass}, we expect to reach a
reconstruction rate of 
\begin{eqnarray}
N_{D^{*+}} = 124.8\cdot 10^3 \;\; (166.4\cdot 10^3)\cdot year^{-1}.
\label{d*rate}
\end{eqnarray}
A rigorous evaluation of acceptance will change these figures, because
COMPASS experiment uses kinematic cuts different from ours. Nevertheless,
the equality (\ref{d*rate}) gives a good idea about the actual
reconstruction rate, attainable in the proposed experiment. }

The observed spin asymmetry $A_{D^{*}}^{exp}$ of $D^{*+}$-meson
photoproduction with account of the beam and target polarizations is given
by 
\begin{eqnarray}
A_{D^{*}}^{exp}=P_b\;P_t\;f_t\;A_{D^{*}}^{\gamma N}.
\end{eqnarray}
Here the asymmetry $A_{D^{*}}^{\gamma N}$ is the ratio of helicity
dependent
and helicity averaged cross sections for $D^{*+}$-meson production in $%
\gamma N$-collisions, calculated with account of kinematic cuts.
Parameters $%
P_b,\;P_t$ are the beam and target polarizations, $f_t$ denotes the
fraction
of the polarized target nucleons. The target characteristics are given in
Tabl. 3, while the gamma beam polarization $P_b$ is about unity.

\vspace*{5mm}

\begin{table}[tbp]
\caption{The target parameters. Here $P_t$ is the target polarization,
$f_t$
denotes the fraction of the polarized target nucleons, $T_n$ stands for
the
target density, and $L$ is the target length.}
\begin{tabular}{|c|c|c|c|c|}
\hline
&  &  &  &  \\ 
& \hspace*{4mm}$f_t$ \hspace*{4mm} & \hspace*{8mm} $P_t$\hspace*{8mm} & $%
T_n\;( 10^{25} cm^{-2})$ & \hspace*{6mm}$L \; (cm)$ \hspace*{6mm} \\
\hline
&  &  &  &  \\ 
butanol & $0.24$ & 0.8 & 4 & 40 \\ \hline
\end{tabular}
\end{table}

\noindent

We impose the $p_t$ cut on $D^{*}$-meson transverse momentum and choose
three values of $p_t$: $0,\;1,\;2\;GeV/c$. The statistical precision of $%
A_{D^{*}}^{\gamma N}$ with account of target properties is given by 
\begin{eqnarray}
\Delta A_{D^{*+}}^{\gamma N}\sim \frac 1{{\sqrt{N_{D^{*+}}}}%
\;f_t\;P_b\;P_t}=2.08\;(1.81)\cdot 10^{-2},  \label{accuracy}
\end{eqnarray}
for the cut $p_t\geq 1\;GeV/c$. One can achieve even better statistical
precision by detecting $D^{*}$ mesons in the total kinematic region of
$p_t$
(Tabl.4). On the other hand, ``strong'' cut on the transverse momentum,
say $p_t\geq 2\;GeV/c$, will lower the statistical accuracy by a factor of
three,
because the corresponding cross section does not exceed 3--4 nb. Our
opinion
is that the ``moderate'' restriction $p_t\geq 1\;GeV/c$ is best suited for
investigation of gluon polarization. Under this cut the total cross
section
is still sizeable, which results in a reasonable value of statistical
precision, moreover it permits to decrease background and reduce
uncertainties in the c-quark fragmentation process.

\begin{table}[tbp]
\caption{The registration rates of $D^{*+}$ mesons at TESLA. Here $\sigma$
is the cross section of inclusive $D^*$ meson production, $N_{D^{*+}}$
denotes the total number of observed $D^{*+}$ mesons per year with account
of efficiency, $\Delta A_{D^{*+}}^{\gamma N}$ stands for the statistical
precision. }
\begin{tabular}{|c|c|c|c|}
\hline
&  &  &  \\ 
& \hspace*{6mm}$\sigma (nb)$ \hspace*{6mm} & \hspace*{8mm}
$N_{D^{*+}}(10^3)$%
\hspace*{8mm} & \hspace*{5mm}$\Delta A_{D^{*+}}^{\gamma N}(10^{-2})$%
\hspace*{5mm} \\ \hline
&  &  &  \\ 
$p_t\geq 0$ & 126 (215) & 524.2 (894.4) & 0.72 (0.55) \\ \hline
&  &  &  \\ 
$p_t\geq 1\; GeV/c$ & 30 (40) & 124.8 (166.4) & 1.47 (1.28) \\ \hline
&  &  &  \\ 
$p_t\geq 2\; GeV/c$ & 4.1 (5) & 17.1 (20.8) & 3.99 (3.61) \\ \hline
\end{tabular}
\end{table}

\vspace*{5mm}

\hspace*{45mm} {\bf 4.2 $D^0$ tagging } \hspace*{20mm}

Experimentally, total number of $D^0$-mesons per charm event is
approximately equal to the sum of $D^{*+}$- and $D^{*0}$-meson production
rates. Produced $D^0$ may be detected via the simplest two - body decay 
\begin{eqnarray}
D^0 \rightarrow K^- \pi^+,  \nonumber
\end{eqnarray}
with the branching ratio of $3.8\%$. An estimate shows that $D^0$- meson
reconstruction rate is more than corresponding quantity for $D^{*+}$-meson
by a factor of three due to a larger value of both the $D^0$-production
cross section and decay branching ratio. However, unlike the $D^{*+}$%
-tagging, a background to the $D^0$-production remains significant even
after account of kinematic cuts, and it exceeds signal, for example at
COMPASS, by a factor of four \cite{Compass}. It means that correct
evaluation of both the asymmetry and statistical precision of measurement
requires a detailed analysis of the background, which is beyond the scope
of
our paper.

\hspace*{35mm} {\bf 4.3 Single muon tagging} \hspace*{20mm}

The most simple method to select charm production events consists in the
detecting of muons from semileptonic $D$-meson decays. Muons coming from
light meson decays, as well as Bethe-Heitler pairs, contribute mainly to
the
low values of $p_t$, so one may reduce these backgrounds by imposing cut
on $%
D$-meson transverse momentum, say $p_t \geq 1\;GeV/c$. Assuming the
branching ratio of the decay 
\begin{eqnarray}
D^{0\;(+)}\rightarrow \mu + X,  \nonumber
\end{eqnarray}
to be $6.8\;\%$ for $D^0$ and $17.2\;\%$ for $D^+$-meson, and ratio of the
$D$%
-meson production rates per charm event to be 
\begin{eqnarray}
\frac{N_{D^0}}{N_{D^+}} \sim 3,  \nonumber
\end{eqnarray}
one expect about $N_{\mu} = 4.51\;(6.02)\cdot 10^6$ prompt muons at TESLA
for $m_c = 1.5,\;( 1.3)\; GeV/c^2$ correspondingly. In average, muon
acquires a transverse momentum equal to about one half of the parent meson
mass, that is $1\; GeV/c$ in our case. To obtain a rough estimate, suppose
that only one half of muons coming from decays of $D$-mesons with $p_t
\geq
1\;GeV/c$ will get a transverse momentum larger than $2\;GeV/c$. In this
case a reconstruction rate is given by 
\begin{eqnarray}
N_{\mu} \sim 2.26\;(3.01)\cdot 10^6,  \nonumber
\end{eqnarray}
under assumption of $100\;\%$ muon detection efficiency. The precision of
the asymmetry measurement with account of the target properties can be
roughly estimated as 
\begin{eqnarray}
\Delta A_{D}^{\gamma N} \sim \frac{1}{{\sqrt {N_{\mu}}}\;f_t\;P_b\;P_t } =
3.5\;(3.0)\cdot 10^{-3}.  \label{mu_accuracy}
\end{eqnarray}
The exact evaluation of the statistical error requires a close
investigation
of the background. Nevertheless Eq. (\ref{mu_accuracy}) gives a good idea
about an actual precision one may hope to reach at TESLA experiment. We
remind, for comparison, that SLAC collaboration expected to obtain
statistical accuracy of asymmetry measurement about $6\cdot10^{-3}$ under
severe conditions of large background coming from pile up of the events.

\section{ Conclusion}

In the present paper we have considered an opportunity to use polarized
real
photon beam for investigation of polarized gluon distributions. Our
estimate
back up the possibility to achieve a higher accuracy in the measurement of
charm photoproduction asymmetry compared with planned experiments at SLAC
and CERN. The proposed experiment may give unambiguous information about
both the total value and the $x$- dependence of gluon polarization, which
will permit to reduce significantly the number of acceptable models
decsribing nucleon spin effects.


\begin{thebibliography}{99}
\bibitem{EMC}  J. Ashman et al, Phys. Lett. {\bf B206} (1988), 364;
\\Nucl.
Phys. {\bf B328} (1989) 1.

\bibitem{Ellis}  J. Ellis and R. Jaffe, Phys. Rev. {\bf D9} (1974) 1444;
\\%
Phys. Rev. {\bf D10} (1974) 1664 (E).

\bibitem{Efremov}  A. Efremov and O. Teryaev, JINR-E2-88-287 and Phys.
Lett. 
{\bf B240} (1990) 200; \\G. Altarelli and G. Ross, Phys. Lett. {\bf B212}
(1988) 391.

\bibitem{MIT}  R. L. Jaffe, Phys. Lett. {\bf B365} (1996) 359.

\bibitem{charmsea}  J.J. Aubert et al, Phys. Lett. {\bf B110} (1982) 73.

\bibitem{Compass}  COMPASS Proposal, CERN/SPSLC 96-14, SPSC/P 297.

\bibitem{asympropos}  S. Alekhin, V. Borodulin and S. Sultansoy,
Int.J.Mod.
Phys. {\bf A8} (1993) 1603; \\S. Atag et.al.,Europhys. Lett. 29 (1995)
815;
Nucl.Instr.and Meth. {\bf A381} (1996) 23 ;\\V. Borodulin et al.,
Proceedings of the 2$^{nd}$ spin workshop, Zeuthen, 1997. Eds.:
J.Blumlein,
A.DeRoeck, T.Gehrmann and W.D.Nowak. p. 439; \\M.Duren, the same
proceedings, p. 414.

\bibitem{tesla}  Conceptual Design of a 500 GeV e+e- Linear Collider with
Integrated X-ray Laser Facility, Eds: R.Brinkmann. G.Materlik, J.Rossbach,
A.Wagner , DESY 1997-048, ECFA 1997-182.

\bibitem{Ginzb}  I. F. Ginzburg et al., Nucl. Instr. and Meth. {\bf 205}
(1983) 477.

\bibitem{Atag}  S. Atag et al., Nucl. Instr. and Meth. {\bf A381} (1996)
21.

\bibitem{mirror}  Ciftci et.al., NIM {\bf A365} (1995) 317.

\bibitem{difcrossec}  G. Altarelli and W.J. Stirling, Particle World {\bf
1}
(1989) 40; \\W. Vogelsang, in Physics at HERA, Vol. 1, eds. W. Buchmuller
and G. Ingelman, (1991) 389.

\bibitem{Gehrmann}  T. Gehrmann and W.J. Stirling, DTP/95/82,
hep-ph/9512406.

\bibitem{Peterson}  C. Peterson et al., Phys. Rev. {\bf D27} (1983) 105.

\bibitem{dis}  S.I. Alekhin, IHEP 96-79, hep-ph/9611293, 1996, to appear
in
Eur. Phys. Jour.

\bibitem{numercode}  T.A. Merkulova and G.G. Takhtamyshev,
JINR-E11-95-255,
1995.

\bibitem{Stratman}  I. Bojak and M. Stratmann, DTP/98/22, hep-ph/9804353;
\\%
I. Bojak and M. Stratmann, DTP/98/36, hep-ph/9807405.
\end{thebibliography}
\end{document}